\documentclass[prd,aps,floats,twocolumn]{revtex4}
\usepackage{amssymb}
\usepackage{amsmath}
\usepackage{mathrsfs}
\usepackage{latexsym}
\usepackage[dvips]{graphicx}

\newcommand{\e}[1]{e^{#1}}

\newcommand{\mathcalr}[1]{{\ensuremath{\mathscr{#1}}}}

\begin{document}

\title{A Toy Model of a Fake Inflation}

\author{M. Novello}
\affiliation{
Institute of Cosmology, Relativity and Astrophysics ICRA/CBPF \\
Rua Dr. Xavier Sigaud 150, Urca 22290-180 Rio de Janeiro,
RJ-Brazil}
\author{E. Huguet and J. Queva}
\affiliation{Laboratoire APC\footnote{UMR 7164 (CNRS, Universit\'e
Paris 7, CEA,
Observatoire de Paris).}, 11 rue M. Berthelot,
F-75251 Paris Cedex 05, France,\\and\\
Universit\'e Paris 7-Denis-Diderot,
bo\^ite 7020, F-75251 Paris Cedex 05, France\footnote{Present address of authors.}.}
\date{\today}

\begin{abstract}

Discontinuities in non linear field theories propagate through
null geodesics in an effective metric that depends on its dynamics
and on the background geometry.  Once information of the geometry
of the universe comes mostly from photons, one should carefully
analyze the effects of possible nonlinearities on Electrodynamics
in the cosmic geometry. Such phenomenon of induced metric is
rather general and may occurs for any nonlinear theory
independently of its spin properties. We limit our analysis here
to the simplest case of non linear scalar field. We show that a
class of theories that have been analyzed in the literature,
having regular configuration in the Minkowski space-time
background is such that the field propagates like free waves in an
effective deSitter geometry. The observation of these waves would
led us to infer, erroneously, that we live in a deSitter universe.

\end{abstract}


\maketitle

\section{Introduction}
We learned from General Relativity that the geometry of spacetime
is guided by gravitational forces. The possibility of such
identification of gravity with geometry relies on the universality
of such interaction. Nevertheless, in certain cases, dealing with a
not so general interaction, it is worth to describe certain kind
of evolutionary processes by appealing to an effective
modification of the geometry. This is the case, for instance, with
the propagation of waves of spin zero (scalarons), spin one
(photons) in nonlinear field theories and the sonic analogue of
Black Holes \cite{NVV}, \cite{Novello1}, \cite{Novello2}. Indeed,
it was shown in these papers that the discontinuities of nonlinear
theories propagate through curves which are null geodesics of an
effective geometry $\widehat{g}_{\mu\nu}$ which depends not only
on the dynamics but on the properties of the background field too.

The importance of such analogue models, dealing with modifications
of the geometry which are not consequences of gravitational
processes, is related to the impossibility to control
gravitational fields in laboratory experiments. The fact that we
can, in principle, produce specific cases of geometries which have
similar properties of solutions of the equations of general
relativity, allows us to understand a little better, the behavior
of matter in gravity interaction by the analysis of analogous
situations, using others interactions, which are capable to be
under our experimental control. The case of the emission of
radiation by a Black Hole is a typical one, once it is understood
\cite{Matt} that a similar behavior could occurs either in sonic
or in electromagnetic Black Holes \cite{Novello2}.

Such an effective description allow us to pose the following
question: is it possible, for a given field theory, to exhibit a
configuration, satisfying nonlinear equations of motion in
Minkowski background and satisfying the property that the
propagation of the waves of the field experiences in this state a
prescribed geometry, to be specific, e.g., the one described by
deSitter?

In this letter we show that the answer is positive and we exhibit an
example that corresponds to a situation in which it occurs. In order
to simplify our calculation we consider the case of a nonlinear
scalar field configuration\footnote{A very important generalization
of the present example should be for nonlinear theories of the
Electromagnetism. We will analyze this case in a future paper}. The
reason for this is twofold: it is the simplest case to deal with and
it constitutes a fundamental element of the scenario that
cosmologists are using nowadays as viable candidates to represent
the basic ingredient of the matter content of the universe, that is,
dark energy. According to this last motivation, our study here can
be understood as a toy model for a fake inflation.

\section{The nonlinear dynamics of a scalar field}
The observation of the acceleration of the universe has brought
into attention new candidates to describe forms of matter with
some unusual properties. One of these is the so-called Chaplygin
gas \cite{Chimento}. A remarkable property of this fluid is that
its energy content can be equivalently described in terms of a
scalar field that satisfies a non linear dynamics obtained from
the Born--Infeld action. A certain number of distinct models of
non linear theories is being studied.  The important point which
is relevant for our analysis concerns the propagation of the
associated scalar waves.

We consider a class of Lagrangians \cite{makler} of the form
$\mathcalr{L}(w, \varphi)=f(w) - V(\varphi)$, where $w :=
\partial_\mu\varphi\partial^\mu\varphi$. The first and second
derivatives of $\mathcalr{L}$ with respect to $w$ are denoted
$L_w$ and $L_{ww}$ respectively. The equation of motion for
$\varphi$ reads:
\begin{equation}\label{EqMotion}
\frac{1}{\sqrt{-g}}\partial_\mu\Bigl(\sqrt{-g} g^{\mu\nu} L_w
(\partial_\nu \varphi)\Bigr) = - \frac{1}{2} \, \frac{\delta
V}{\delta\varphi}.
\end{equation}
We are interested to evaluate the characteristic surfaces of wave
propagation of this theory. The most direct and elegant way to
pursue this goal is to use the Hadamard formalism \cite{Hadamard}.
Let $\Sigma$ be a surface of discontinuity of the scalar field
$\varphi.$ We define the discontinuity of an arbitrary function $f$
to be given by:
\begin{equation}\label{DiscDef}
\left[ f(x) \right]\Bigl\vert_\Sigma = \lim_{\epsilon \to 0^+}
\bigl( f(x + \epsilon) - f(x - \epsilon)\bigr)
\end{equation}
We take that $\varphi$ and its first derivative $\partial_\mu
\varphi$ are continuous across $\Sigma$, while the second
derivatives presents a discontinuity:
\begin{eqnarray}\label{CondHadamard}
\left[ \varphi(x) \right]\Bigl\vert_\Sigma &=& 0,\\
\left[ \partial_\mu \varphi(x) \right]\Bigl\vert_\Sigma &=& 0,\\
\left[ \partial_\mu\partial_\nu\varphi(x) \right]\Bigl\vert_\Sigma
&=& k_\mu k_\nu \xi(x),
\end{eqnarray}
where $ k_\mu := \partial_\mu \Sigma$ is the propagation vector
and $\xi(x)$ the amplitude of the discontinuity. Once
${\displaystyle \frac{\delta V}{\delta\varphi}}$ is continuous
across $\Sigma$ and applying (\ref{CondHadamard}) to
(\ref{EqMotion}) we find:
\begin{equation*}
k_\mu k_\nu \Bigl(L_w g^{\mu\nu} + 2 L_{ww}
\partial^\mu\varphi\partial^\nu\varphi\Bigr) = 0.
\end{equation*}
This equation suggests the introduction of an effective metric
defined by:
\begin{equation}\label{gEff}
\widehat{g}^{\mu\nu} := L_w g^{\mu\nu} + 2 L_{ww}
\partial^\mu\varphi\partial^\nu\varphi
\end{equation}
The inverse $\widehat{g}_{\mu\nu}$ of (\ref{gEff}) is obtained by
using the ansatz $\widehat{g}_{\mu\nu} = A g_{\mu\nu} +
B\partial_\mu\varphi\partial_\nu\varphi$ where the unknown
coefficients $A$ and $B$ are determined through the condition
$\hat{g}^{\mu\alpha}\hat{g}_{\alpha\nu} = \delta^\mu_\nu$. This
leads to:
 \begin{equation}\label{gEffInv}
\widehat{g}_{\mu\nu} := \frac{1}{L_w}\Bigl(g_{\mu\nu} - \frac{2
L_{ww}}{\Psi}
\partial_\mu\varphi\partial_\nu\varphi\Bigr),
\end{equation}
where we defined $\Psi := L_w + 2 w L_{ww}$.

\section{Methodology} Since the scalar field "see" the effective
geometry one can ask for non linear Lagrangians leading to a given
effective geometry in a fixed background. To this end one proceeds
by choosing a Lagrangian, determining the corresponding effective
geometry and solving the Euler-Lagrange equations for $\varphi$.
Unfortunately since the effective metric depends on the field,
such an approach is often intractable. A convenient means to
simplify the problem is to choose $\Psi$, this allows to partly
control the interplay between the Lagrangian and the effective
geometry (\ref{gEffInv}). Let us examine the simplest\footnote{The
background Minkowski geometry is given in the standard gaussian
coordinate system. 
We note that the conditions $\Psi = 1$ and $\varphi = \varphi(t)$ implies
that the variable $t$ is the global time for the effective metric.} case
where $\Psi = 1$ which we use hereafter. This choice obviously
simplify (\ref{gEffInv}) and is equivalent to the equation: $ L_w
+ 2 w L_{ww} = 1$. This equation can be straightforwardly
integrated to yield:
\begin{equation}
\mathcalr{L} = w + 2 \lambda \sqrt{w} + C,
\label{5Nov2}
\end{equation}
where $\lambda$ is a non-zero c-number and $C$ a constant with
respect to $w$, in particular one can set $C = - V(\varphi)$. This
case is worth of considering in particular because of the
properties it provides for the effective metric, but besides this
it can be understood as a perturbation of the standard linear
theory. Just to present a toy model that corresponds to a specific
"fake inflation" we will restrict  the case in which the potential
take the form\footnote{See for instance, F. Finelli and R.
Branderberger, arXiv:hep-th/0112249, November 2002.}
\begin{equation}
 V(\varphi) = - \lambda^2 \mathbf{x} \,
(1 + \frac{\mathbf{x}}{2}), \label{5Nov1}
\end{equation}
 where we have defined $\mathbf{x}
\equiv \e{-\frac{2 H}{\lambda} \varphi}.$

\section{Effective FRW metric
in a Minkowskian background}

We first set the background metric to the Minkowski metric
$\eta_{\mu\nu}$, and use the convention $(+,-,-,-)$. We now show
that for a field $\varphi$ depending only on time $\varphi =
 \varphi(t)$ and satisfying the Lagrangian (\ref{5Nov2}) the effective
metric $\widehat{g}$ felt by $\varphi$ is a spatially flat FRW
metric (with usual notations):
\begin{equation}\label{FRW}
ds^2 = dt^2 - a^2(t) \bigl(dr^2 + r^2 d\Omega^2\bigr).
\end{equation}

Since $\varphi$ do not depends on spatial coordinates the
Euler-Lagrange equations reduce simply to:
\begin{equation*}
\ddot{\varphi}\bigl(L_w + 2 (\dot{\varphi})^2 L_{ww}\bigr) = -
\frac{1}{2} \frac{\delta V}{\delta\varphi},
\end{equation*}
where a dot means a derivative with respect to the time.  At this
point we remark that since $w = (\dot{\varphi})^2$ the above
equation reads:
\begin{equation}\label{PhiPsiV}
\ddot{\varphi}\Psi = - \frac{1}{2} \frac{\delta V}{\delta\varphi}.
\end{equation}
Now, the effective invariant length element reads:
\begin{equation}\label{ds2}
ds^2 = \widehat{g}_{\mu\nu}dx^\mu dx^\nu =  dt^2 -
\frac{1}{L_w}\Bigl(dr^2 + r^2 d\Omega^2\Bigr).
\end{equation}
Note that $\Psi = 1$, that is $2L_{ww} = 1-L_w$, leads to
$\widehat{g}_{tt} = 1$. Let us set the expansion factor on the
effective FRW geometry to an inflationary form: $a(t) = \e{H t}$,
$H$ being a reel positive parameter. For that choice the equation:
\begin{equation*}
a(t) = \frac{1}{L_w},
\end{equation*}
leads to:
\begin{equation}\label{EqPhiInfl}
\sqrt{w} = \frac{\lambda}{\e{-2 H t} - 1}.
\end{equation}
Since $\sqrt{w}$ is positive $\lambda$ must be negative.  Assuming
$\dot{\varphi} \leqslant 0$ (calculations for $\dot{\varphi}
\geqslant 0$ are analogous) the above equation can be integrated
to~:
\begin{equation}\label{PhiIfl}
\varphi = \frac{\lambda}{2 H} \ln (\e{2 H t} - 1) + K,
\end{equation}
where $K$ is a constant, which we set equal to zero. Solving
(\ref{PhiIfl}) for $t$ allows to integrate (\ref{PhiPsiV}) to
obtain precisely the form exhibited in equation (\ref{5Nov1}) of
the potential. In other words, observation of the effective
geometry $\widehat{g}_{\mu\nu}$ would led us to believe,
erroneously, that we live in a deSitter geometry. Although we are
dealing here with a toy model, a similar situation can occur for
other non linear theories, like nonlinear Electrodynamics.

\section*{Acknowledgments}
M Novello acknowledges support of FAPERJ and CNPq and Universit\'e
de Paris 7.

\end{document}